\documentclass[a4paper]{spie}
\usepackage[]{graphicx}
\newcommand{\bra}[1]{\mbox{$\langle{#1}|$}}
\newcommand{\ket}[1]{\mbox{$|{#1}\rangle$}}

\title{Geometric information in eight dimensions vs. quantum information}
\author{Victor I. Tarkhanov\supit{a} and Michael M. Nesterov\supit{b}
\skiplinehalf
\supit{a}St. Petersburg State Polytechnic University, St. Petersburg, Russia \\
\supit{b}St. Petersburg Institute for Informatics and Automation, Russian Academy of Sciences, St. Petersburg, Russia}

\authorinfo{Further author information: (Send correspondence to
V.I.T.)\\
V.I.T.: E-mail: vtarkhanov@mail.ru \\
M.M.N.: E-mail: nesterov@epr.pu.ru}

\begin{document}
\maketitle

%%%%%%%%%%%%%%%%%%%%%%%%%%%%%%%%%%%%%%%%%%%%%%%%%%%%%%
\begin{abstract}
Complementary idempotent paravectors and their ordered compositions, are used to represent multivector basis elements of geometric Clifford algebra $\mathcal{G}_{3,0}$ as the states of a geometric byte in a given frame of reference. Two layers of information, available in real numbers, are distinguished. The first layer is a continuous one. It is used to identify spatial orientations of similar geometric objects in the same computational basis. The second layer is a binary one. It is used to manipulate with 8D structure elements inside the computational basis itself. An oriented unit cube representation, rather than a matrix one, is used to visualize an inner structure of basis multivectors. Both layers of information are used to describe unitary operations --- reflections and rotations --- in Euclidian and Hilbert spaces. The results are compared with ones for quantum gates. Some consequences for quantum and classical information technologies are discussed.
\end{abstract}

%%%%%%%%%%%%%%%%%%%%%%%%%%%%%%%%%%%%%%%%%%%%%%%%%%%%%%%%%%%%%%
% воскресенье, ќкт€брь 28, 2007  12:06

\section{INTRODUCTION}
\label{sect:intro}

The idea to use geometric Clifford algebras in information processing technologies is not new. It was used, for example, by T.\,F. Havel et al.\cite{Havel_2000} to replace quantum mechanics in quantum NMR computing. Along with obvious advantages of the method, it was mentioned, that geometric algebra had a lot of extra degrees of freedom. Special correlators were invented to eliminate them by reducing all imaginary units to a single one.

More radical idea was introduced by D. Hestenes \cite{Hestenes_86}, who suggested to use geometric algebras as a unified language to combine all existing theories as in mathematics and in physics. On this way he had got a success \cite{Hestenes_02}, but all the efforts were directed on those parts of geometric algebras which were in common with one or more of these theories. Again those parts and features of geometric algebras, which were beyond the existing theories were laid aside.

It was silently admitted, that geometric algebras could be used as a universal tool to describe all existing theories, but they would hardly bring anything new into them.
Is it really so?

Geometric algebra is a synthesis of two calcules: algebra and geometry. It's main idea is to deal with geometric objects in terms of algebra in a given vector space. But there are not only vectors and scalars among its elements. There are objects of other grades and clusters of mixed grades as well. Some of them are stable ones, like fullerenes in physics. Not all of them are suitable for measurements, especially those with rather special or extraordinary algebraic and geometric properties. But they are sound algebraic elements which could be used to describe physics involved into information technologies.

We concentrate here on using 4D and 8D clusters of geometric algebra $\mathcal{G}_{3,0}$ --- complementary idempotent paravectors and their ordered products --- to show the way to get some new ideas in classical and quantum information technologies.

\section{3D EUCLIDEAN VECTOR SPACE IN EIGHT DIMENSIONS}
\label{3Din8D}

We begin with geometric algebra $\mathcal{G}_{3,0}$, which is spanned on vectors of 3D Euclidean space and has an 8D multivector basis \cite{tar_02}:
\begin{equation}\label{1}
    \{e_0,e_1,e_2,e_3,e_{12},e_{23},e_{13},e_{123}\},
\end{equation}
where $e_{ik}=e_ie_k$, $i,k = 1,2,3$, and $e_{123} = e_1e_2e_3$. There is a unit scalar (grade 0), $e_0$; three unit vectors (grade 1), $e_1,e_2,e_3$; three unit bivectors (grade 2), $e_{12},e_{13},e_{23}$, and a unit trivector (grade 3), $e_{123}$, among them.

Inside this basis we combine a unit scalar $e_0$ and each of three basis vectors, $e_i$, $i=1,2,3$, into clusters of likeness and distinction: positive and negative idempotent paravectors, $P_i = \frac{1}{2}(e_0 + e_i)$ and $N_i = \frac{1}{2}(e_0 - e_i)$, respectively \cite{tar_06}.

They are sound 4D objects, hermitian and invariant to multiplication by themselves, which are normed to \emph{themselves}, not to single scalars. Each pair of them is orthogonal (linear independent), $P_iN_i = N_iP_i = 0$, and complementary in two senses: a scalar one, $P_i+N_i=e_0$, and a vector one, $P_i-N_i=e_i$.

These complementarities are used to define geometric bits
\begin{eqnarray}
\nonumber    P_1 + N_1 = e_0; \qquad P_1 - N_1 = e_1; \qquad \{+,-\}; \\
    P_2 + N_2 = e_0; \qquad P_2 - N_2 = e_2; \qquad \{+,-\}; \label{2} \\
\nonumber    P_3 + N_3 = e_0; \qquad P_3 - N_3 = e_3; \qquad \{+,-\}.
\end{eqnarray}
There is a common ``ground'' scalar state, $e_0$, for all of them, and different ``excited'' vector states depending on spatial properties of unit vectors, used in their inner structure compositions.

These geometric bits are combined into a geometric byte with a sort of binomial formula, $$\mathcal{A} = (P_1\pm N_1)(P_2\pm N_2)(P_3\pm N_3),$$ to get basis multivectors, $e_0$, $e_1$, $e_2$, $e_3$, $e_{12}$, $e_{13}$, $e_{23}$, $e_{123}$, as the states of the byte \cite{tar_06}:
\begin{eqnarray}
% \nonumber to remove numbering (before each equation)
\nonumber  e_0 &=& (P_1 + N_1)(P_2 + N_2)(P_3 + N_3); \qquad \{+,+,+\} \\
\nonumber  e_1 &=& (P_1 - N_1)(P_2 + N_2)(P_3 + N_3); \qquad \{-,+,+\} \\
\nonumber  e_2 &=& (P_1 + N_1)(P_2 - N_2)(P_3 + N_3); \qquad \{+,-,+\} \\
e_3 &=& (P_1 + N_1)(P_2 + N_2)(P_3 - N_3); \qquad \{+,+,-\} \label{3} \\
\nonumber  e_{12} &=& (P_1 - N_1)(P_2 - N_2)(P_3 + N_3); \qquad \{-,-,+\} \\
\nonumber  e_{23} &=& (P_1 + N_1)(P_2 - N_2)(P_3 - N_3); \qquad \{+,-,-\} \\
\nonumber  e_{13} &=& (P_1 - N_1)(P_2 + N_2)(P_3 - N_3); \qquad \{-,+,-\} \\
\nonumber  e_{123} &=& (P_1 - N_1)(P_2 - N_2)(P_3 - N_3); \qquad \{-,-,-\}.
\end{eqnarray}

\begin{figure}[h]
 \centering \includegraphics[height=3cm]{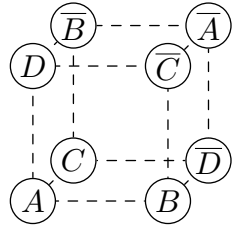}
 \caption{A unit cube representation.}\label{f:1}
 \end{figure}

Opening the brackets brings us to a set of binary coded superpositions of the same 8D structure elements for all basis multivectors, which have no analogs in quantum or classical theories. Each of them is a scaled superposition of all basis multivectors and vise versa:
\begin{eqnarray}
% \nonumber to remove numbering (before each equation)
\nonumber  A &=& P_1P_2P_3 = \frac{1}{8}(e_0+e_1+e_2+e_3+e_{12}+e_{13}+e_{23}+e_{123}); \\
\nonumber  B &=& N_1P_2P_3 = \frac{1}{8}(e_0-e_1+e_2+e_3-e_{12}-e_{13}+e_{23}-e_{123}); \\
\nonumber  C &=& P_1N_2P_3 = \frac{1}{8}(e_0+e_1-e_2+e_3-e_{12}+e_{13}-e_{23}-e_{123}); \\
  D &=& P_1P_2N_3 = \frac{1}{8}(e_0+e_1+e_2-e_3+e_{12}-e_{13}-e_{23}-e_{123}); \label{4} \\
\nonumber  \overline{D} &=& N_1N_2P_3 = \frac{1}{8}(e_0-e_1-e_2+e_3+e_{12}-e_{13}-e_{23}+e_{123}); \\
\nonumber  \overline{C} &=& N_1P_2N_3 = \frac{1}{8}(e_0-e_1+e_2-e_3-e_{12}+e_{13}-e_{23}+e_{123}); \\
\nonumber  \overline{B} &=& P_1N_2N_3 = \frac{1}{8}(e_0+e_1-e_2-e_3-e_{12}-e_{13}+e_{23}+e_{123}); \\
\nonumber  \overline{A} &=& N_1N_2N_3 = \frac{1}{8}(e_0-e_1-e_2-e_3+e_{12}+e_{13}+e_{23}-e_{123}).
\end{eqnarray}
They are associated \cite{tar_06} with the oriented octants of a unit cube and are labeled with a letter in each vertex\footnote{An overline means a conjugate operation of space reversion, in which all vectors change their signs.}, Fig.\ref{f:1}.
They are the same oriented octants as in a Cartesian frame of reference, but scaled and shifted from the origin to compose a unit cube form.

In these notations complementary basis paravectors are the opposite sides of the unit cube:
\begin{eqnarray}
% \nonumber to remove numbering (before each equation)
\nonumber  P_1 &=& A + \overline{B} + C + D = \includegraphics[height=1cm]{vers.2}; \qquad N_1 = \overline{A} + B + \overline{C} + \overline{D} =
\includegraphics[height=1cm]{vers.3}; \\
  P_2 &=& A + B + \overline{C} + D =
\includegraphics[height=.85cm]{vers.4}; \qquad
  N_2 = \overline{A} + \overline{B} + C + \overline{D} =
\includegraphics[height=.85cm]{vers.5}, \label{5}\\
\nonumber  P_3 &=& A + B + C + \overline{D} =
\includegraphics[height=.85cm]{vers.6}; \qquad
  N_3 = \overline{A} + \overline{B} + \overline{C} + D =
\includegraphics[height=.85cm]{vers.7}.
\end{eqnarray}

The inner structure of basis multivectors in these notations is coded as
\begin{eqnarray}
% \nonumber to remove numbering (before each equation)
\nonumber  e_0 &=& (A+\overline{A}) + (B+\overline{B}) +
(C+\overline{C}) + (D+\overline{D}) = \includegraphics[height=1cm]{vers.8}; \\
\nonumber  e_1 &=& (A-\overline{A}) - (B-\overline{B}) +
(C-\overline{C}) + (D-\overline{D}) = \includegraphics[height=1cm]{vers.9}; \\
\nonumber  e_2 &=& (A-\overline{A}) + (B-\overline{B}) -
(C-\overline{C}) + (D-\overline{D}) = \includegraphics[height=1cm]{vers.10}; \\
\nonumber  e_3 &=& (A-\overline{A}) + (B-\overline{B}) +
(C-\overline{C}) - (D-\overline{D}) = \includegraphics[height=1cm]{vers.11}; \\
  e_{12} &=& (A+\overline{A}) - (B+\overline{B}) - (C+\overline{C}) +
  (D+\overline{D}) =   \includegraphics[height=1cm]{vers.12}; \label{6} \\
\nonumber e_{23} &=& (A+\overline{A}) + (B+\overline{B}) - (C+\overline{C}) -
(D+\overline{D}) = \includegraphics[height=1cm]{vers.13}; \\
\nonumber  e_{13} &=& (A+\overline{A}) - (B+\overline{B}) + (C+\overline{C}) -
(D+\overline{D}) = \includegraphics[height=1cm]{vers.14}; \\
\nonumber  e_{123} &=& (A-\overline{A}) - (B-\overline{B}) -
(C-\overline{C}) - (D-\overline{D}) =
\includegraphics[height=1cm]{vers.15}.
\end{eqnarray}
Here green circles correspond to structure elements taken with the sign ``plus'', and blue circles --- to ones, taken with the sign ``minus''. So there are sets of binary coefficients $+1$ and $-1$ to describe the inner structure of all basis multivectors in terms of the same set of 8D structure elements (\ref{4}).

Note, that each change of state in one of geometric bits (\ref{2}) is coded by 4 changes of sign in structure elements (\ref{4}) and in their superpositions --- basis multivectors (\ref{6}). This is a sort of error correction algorithm intrinsically provided by the laws of algebra.

\section{TWO LAYERS OF INFORMATION}

There are two distinct layers of information, available in real numbers, inside the 8D algebraic space of $\mathcal{G}_{3,0}$ geometric algebra.

The first one is used to compare and identify spatial orientations of similar geometric objects. Its information carriers are real coefficients (coordinates) of basis Clifford numbers, showing their role in a particular spatial decomposition of the object. In this case basis Clifford numbers are the same, and all information about the object is contained in their real coefficients. These may be any real numbers.

When an object is rotated, some of its coordinates are changed in a harmonic or a continuous manner. Being measured in time, their successive changes of values can be treated as flows of analog or digital information.

The second layer is used to identify basis multivectors and their clusters algebraically and to manipulate with their structure elements (\ref{4}). In a unit cube representation its information carriers are real binary numbers, $+1$ and $-1$.

Any rearrangement of structure elements (\ref{4}) inside basis Clifford numbers changes the sense of the computation basis and its elements, their images inside the second information layer, the dimension of space, in which these elements operate, and their properties in algebraic relations. This, in turn, changes the image of the object inside the first information layer.

Let us show it on two simple examples.

\subsection{Unit vector representations}

In a given Cartesian frame of reference $\{e_1,e_2,e_3\}$ any unit vector $a$ is described as
\begin{equation}\label{7}
    a = a_1e_1 + a_2e_2 + a_3e_3.
\end{equation}
Here unit vectors $e_1$, $e_2$ and $e_3$ are fixed, and all information about a particular space orientation of vector $a$ is contained in the set of direction cosines $(a_1,a_2,a_3)$. This is the first layer of information. As there are only three items in the sum and each item is a vector itself, we treat equation (\ref{7}) as a 3D representation (image) of vector $a$.

In geometric algebra $\mathcal{G}_{3,0}$ basis unit vectors $e_1$, $e_2$, $e_3$ are 8D Clifford numbers. So $a$ can be rewritten in the form
\begin{equation}\label{8}
    a = a_1\,\includegraphics[height=1cm]{vers.9} +
    a_2\,\includegraphics[height=1cm]{vers.10} +
    a_3\,\includegraphics[height=1cm]{vers.11}.
\end{equation}
Only three of eight basis multivectors are used here for the decomposition. The inner structure of these Clifford numbers, coded in terms of structure elements (\ref{4}) helps us to identify them as unit vectors in 8D space. This is the second layer of information. It does not change when vector $a$ rotates. The information about the orientation of vector $a$ is contained in the same set of direction cosines $(a_1,a_2,a_3)$.

After combining structure elements (\ref{4}) with their coefficients along the main diagonals of the cube this expression can be rewritten as
\begin{eqnarray}
\nonumber    a &=& (a_1 + a_2 + a_3)\,
\includegraphics[height=1cm]{vers.16} +
    (-a_1 + a_2 + a_3)\,\includegraphics[height=1cm]{vers.17} + \\
    &+& (a_1 - a_2 + a_3)\,\includegraphics[height=.8cm]{vers.18} +
    (a_1 + a_2 - a_3)\,\includegraphics[height=.8cm]{vers.19} = \label{9} \\
\nonumber    &=& (a_1 + a_2 + a_3)(A-\overline{A}) +
    (-a_1 + a_2 + a_3)(B-\overline{B}) +\\
\nonumber    &+& (a_1 - a_2 + a_3)(C-\overline{C}) +
    (a_1 + a_2 - a_3)(D-\overline{D}).
\end{eqnarray}
Now there are four items in the sum, so it is a 4D image of vector $a$. It resembles a sort of a Radon transform. We have changed 3D unit vectors onto some 4D clusters, that is we have increased the dimensionality of basis Clifford numbers. Now they are the inhabitants of a Minkowski space with $\{+,+,+,-\}$ metrics:
\begin{equation}\label{10}
% \nonumber to remove numbering (before each equation)
\left\{ \begin{array}{c}
(A - \overline{A}) = \frac{1}{4}(e_1 + e_2 + e_3 + e_{123}); \\[2ex]
(B - \overline{B}) = \frac{1}{4}(-e_1 + e_2 + e_3 - e_{123}); \\[2ex]
(C - \overline{C}) = \frac{1}{4}(e_1 - e_2 + e_3 - e_{123}); \\[2ex]
(D - \overline{D}) = \frac{1}{4}(e_1 + e_2 - e_3 - e_{123}). \end{array}
\right.
\end{equation}
They are not perpendicular to each other. They are rather orthogonal in the sense of linear independence in 4D space.

We have changed information of the second layer. What has happened with information of the first layer?
It has changed its image, but has preserved all its ingredients --- the direction cosines, $a_1$, $a_2$, $a_3$. Now they are rearranged into some superpositions, but they are again just real numbers. Hence the information of the first layer is the same.

\subsection{Unitary quaternion representations}

Quaternions are 4D even Clifford numbers containing a scalar and a bivector inside \cite{tar_02}. Unitary quaternions are used to describe rotations. In matrix representation a unitary contravariant quaternion usually has a form
\begin{equation}\label{11}
    Q = \alpha P_3 + \beta (e_1P_3) -\beta^*(e_1N_3) + \alpha^* N_3 =
    \left[\begin{array}{cc}
      \alpha & -\beta^* \\
      \beta & \alpha^* \\
    \end{array}\right].
\end{equation}
Here $P_3$, $e_1P_3$, $e_1N_3$ and $N_3$ are paravector names (labels) of matrix elements in geometric algebra $\mathcal{G}_{3,0}$. Complex numbers $\alpha$, $\beta$, $-\beta^*$ and $\alpha^*$, meeting a condition $\alpha\alpha^* + \beta\beta^* = 1$, are called Cayley--Klein parameters.

For a simple positive (anticlockwise) rotation around a unit axis $c=c_1e_1 + c_2e_2 + c_3e_3$ through an angle $\vartheta$ the quaternion (\ref{11}) has a form
\begin{equation}\label{12}
    Q = \exp{(-ic\frac{\vartheta}{2})} = \cos{\frac{\vartheta}{2}} - ic\sin{\frac{\vartheta}{2}}
\end{equation}
with Cayley--Klein parameters
\begin{equation}\label{13}
\left\{ \begin{array}{c}
    \alpha = \cos{\frac{\vartheta}{2}} - ic_3\sin{\frac{\vartheta}{2}}; \\ [2ex]
    \beta = -i(c_1 + ic_2)\sin{\frac{\vartheta}{2}}. \end{array} \right.
\end{equation}
They are complex numbers, complementary in the sense, that, in any given frame of reference, the pare of them contains all information about the angle $\vartheta$ and the axis ($c_1,c_2,c_3$) for the described simple rotation.

In terms of structure elements (\ref{4}), one can write (\ref{11}) as
\begin{equation}\label{14}
    Q(a) = \alpha\,\includegraphics[height=1cm]{vers.6} +
    \beta\,\includegraphics[height=1cm]{vers.20} -
    \beta^*\,\includegraphics[height=1cm]{vers.21} +
    \alpha^*\,\includegraphics[height=1cm]{vers.7},
\end{equation}

For $\alpha = \rho -i\nu$ and $\beta = -i(\mu + i\lambda)$ eq. (\ref{14}) can be rewritten with only real coefficients inside the first information layer. There are two ways to do it.

The first way is a rather traditional one:
%in accordance with Hamilton's definition of quaternions \cite{Hamilton}:
\begin{equation}\label{15}
\nonumber    Q(a) = \rho e_0 - \nu e_{12} - \mu e_{13} - \lambda e_{23} =
    \rho \, \includegraphics[height=1cm]{vers.8} -
    \nu \, \includegraphics[height=1cm]{vers.12} -
    \mu \, \includegraphics[height=1cm]{vers.13} -
    \lambda \, \includegraphics[height=1cm]{vers.14},
\end{equation}
where real numbers $\rho$, $\nu$, $\mu$ and $\lambda$, meeting a condition $\rho^2 + \nu^2 + \mu^2 + \lambda^2 = 1$, are called Euler--Rodrigues parameters. It is built in the computational basis, $\{e_0, e_{12}, e_{13}, e_{23}\}$, which is a basis for a Minkowski space with $\{+,-,-,-\}$ metrics.

The second way is a less obvious one. It is obtained by rearranging structure elements (\ref{4}) with their signs (information of the second layer) and their real coefficients (information of the first layer) into another computational basis:
\begin{eqnarray}
\nonumber    Q(a)&=& (\rho - \nu - \mu - \lambda) \,
    \includegraphics[height=1cm]{vers.22} +
    (\rho + \nu + \mu - \lambda) \,
    \includegraphics[height=1cm]{vers.23} + \\
\nonumber    &+& (\rho + \nu - \mu + \lambda) \,
    \includegraphics[height=.8cm]{vers.24} +
    (\rho - \nu + \mu + \lambda) \,
    \includegraphics[height=.8cm]{vers.25} = \\
    &=& (\rho - \nu - \mu - \lambda)(A+\overline{A}) +
    (\rho + \nu + \mu - \lambda)(B+\overline{B}) + \label{16}\\
\nonumber    &+&(\rho + \nu - \mu + \lambda)(C+\overline{C}) +
    (\rho - \nu + \mu + \lambda)(D+\overline{D}).
\end{eqnarray}
The new basis, $\{(A+\overline{A}), (B+\overline{B}), (C+\overline{C}), (D+\overline{D})\}$, consists of four pairs of structure elements (\ref{4}) lying again along the main diagonals of the unit cube. But now they are combined with plus signs. They are inhabitants of the same 4D Minfowski space with $\{+,-,-,-\}$ metrics:
\begin{equation}\label{17}
\left\{ \begin{array}{c}
(A + \overline{A}) = \frac{1}{4}(e_0 + e_{12} + e_{23} + e_{13}); \\ [2ex]
(B + \overline{B}) = \frac{1}{4}(e_0 - e_{12} - e_{23} + e_{13}); \\ [2ex]
(C + \overline{C}) = \frac{1}{4}(e_0 - e_{12} - e_{23} + e_{13}); \\ [2ex]
(D + \overline{D}) = \frac{1}{4}(e_0 + e_{12} - e_{23} - e_{13}). \end{array}
\right.
\end{equation}
And again some rearrangements of structure elements image inside the second information layer changes the image of the object inside the first information layer.  The new image has the same information, because new coefficients inside the first layer contain all ingredients of the previous image.

\section{PROJECTIONS}

Usually it is a good idea to simplify description of multidimensional geometric objects and their behavior by projecting them onto spaces of less dimensions. There are two ways to project them in our 8D algebraic space: Cartesian and Hilbert ones.

The first way is typical for a 3D vector space, and is widely used in engineering.  Its extension onto 8D multivector space of geometric algebra $\mathcal{G}_{3,0}$ is trivial except for the loss of commutativity in multiplication.  The result is the reduction of the overall 8D space to three mutually orthogonal 4D subspaces.  These projections are used to get orthogonal decompositions inside the first information layer.  The second information layer remains unchanged: all basis elements are in multivector form (\ref{6}).

The second way is a new one. It is based on specific projective properties of 4D clusters --- idempotent paravectors, --- which are in the focus of this article.  In these projections information inside the first layer is invariant, whereas all orthogonal decompositions are made inside the second informamtion layer. Let us consider it in more details.

\subsection{Complementary paravectors for a unit vector}

As it was stated above, for any given unit vector $a$ and the only unit scalar $e_0$ one can built the only cluster of likeness, $P(a) = \frac{1}{2}(e_0 + a)$, and the only cluster of distinction, $N(a) = \frac{1}{2}(e_0 - a)$. They are complementary idempotent paravectors. The first one, $P(a)$, is called positive, because its Cartesian projection onto vector $a$ is the same $P(a)$ with ``plus'' sign:
\begin{equation}\label{18}
    aP(a) = P(a)a = P(a).
\end{equation}
The second paravector, $N(a)$, is called negative, because its Cartesian projection onto vector $a$ is the same $N(a)$ up to the sign. The sign is ``minus'':
\begin{equation}\label{19}
    aN(a) = N(a)a = - N(a).
\end{equation}
These properties can be transferred onto all other elements of the algebra through one-sided multiplications. This is the way to project them, into so called \emph{spinor ideals} of Hilbert spaces, and, hence, this is the reason for a coined name.

There are two orthogonal Hilbert spaces --- positive one, and negative one --- for each unit vector $a$ in a given frame of reference.  They have no common points, so they are orthogonal rather in the sense of parallelism, than in the sense of perpendicularity. And there are two spinor ideals --- a contravariant one, and a covariant one, --- intersecting through a body of two-sided spinors, in each of them.

\subsection{Contravariant Hilbert projections of basis multivectors}

Let us put for simplicity $a = e_3$. Then $P(a) = P_3$, and $N(a) = N_3$. This is the only case when complementary idempotent paravectors are associated with single matrix elements.  We shall use it to compare our results with their analogs in quantum information theory \cite{Nielsen_2002}.  In the same frame of reference all the other vectors are decomposed into complementary idempotent paravectors, which are spread over all four matrix elements simultaneously, as in density martixes, and lose the illusion of 1D objects.

To project basis multivectors (\ref{6}) into an ideal of positive contravariant spinors it is enough to multiply them from the right side by $P_3$:
\begin{eqnarray}
% \nonumber to remove numbering (before each equation)
\nonumber  e_0P_3 &=& e_3P_3 = \frac{1}{2}(e_0 + e_3) = P_3 =
\left[\begin{array}{cc}
      1 & 0 \\
      0 & 0 \\
    \end{array}\right] = \includegraphics[height=1cm]{vers.6}; \\
\nonumber  e_1P_3 &=& e_{13}P_3 = \frac{1}{2}(e_1 + e_{13}) = (e_1P_3) =
\left[\begin{array}{cc}
               0 & 0 \\
               1 & 0 \\
       \end{array}\right] = \includegraphics[height=1cm]{vers.20}; \\
  e_2P_3 &=& e_{23}P_3 = \frac{1}{2}(e_2 + e_{23}) = i(e_1P_3) =
\left[\begin{array}{cc}
      i & 0 \\
      0 & 0 \\
    \end{array}\right] = \includegraphics[height=1cm]{vers.26}; \label{20}\\
\nonumber  e_{12}P_3 &=& e_{123}P_3 = \frac{1}{2}(e_{12} + e_{123}) = iP_3 =
\left[\begin{array}{cc}
               0 & 0 \\
               i & 0 \\
       \end{array}\right] = \includegraphics[height=1cm]{vers.27}.
\end{eqnarray}
There is a two-fold degeneracy in these projections.  From a single projection one cannot say for sure, which of two multivectors was projected to get a particular image, a particular spinor shadow in the positive Hilbert space.

In matrix representation this spinor ideal has a basis consisting of two matrix elements in the left column, $\{P_3,(e_1P_3)\}$. Eight basis multivectors are projected onto them with real or imaginary coefficients.

In the unit cube representation there are four basis paravector clusters with real coefficients and with the same two-fold degeneracy in projections.  Only bottom side structure elements are used here.  So it is a projection of basis multivectors (\ref{6}) to the bottom side of the cube.

To project the same basis multivectors (\ref{6}) into an ideal of negative contravariant spinors, it is enough to multiply them from the right side by $N_3$:
\begin{eqnarray}
% \nonumber to remove numbering (before each equation)
\nonumber  e_0N_3 &=& -e_3N_3 = \frac{1}{2}(e_0 - e_3) = N_3 = \left[\begin{array}{cc}
      0 & 0 \\
      0 & 1 \\
    \end{array}\right] = \includegraphics[height=1cm]{vers.7}; \\
\nonumber  e_1N_3 &=& -e_{13}N_3 = \frac{1}{2}(e_1 - e_{13}) = (e_1N_3) =
\left[\begin{array}{cc}
               0 & 1 \\
               0 & 0 \\
       \end{array}\right] = \includegraphics[height=1cm]{vers.21}; \\
  -e_2N_3 &=& e_{23}N_3 = -\frac{1}{2}(e_2 - e_{23}) = i(e_1N_3)=\left[\begin{array}{cc}
      0 & i \\
      0 & 0 \\
    \end{array}\right] = \includegraphics[height=1cm]{vers.28}; \label{21}\\
\nonumber  -e_{12}N_3 &=& e_{123}N_3 = -\frac{1}{2}(e_{12} - e_{123}) = iN_3 =
\left[\begin{array}{cc}
               0 & 0 \\
               0 & i \\
       \end{array}\right] = \includegraphics[height=1cm]{vers.29}.
\end{eqnarray}
These are Hilbert projections of basis multivectors (\ref{6}) to the upper side of the cube.

In matrix representation this spinor ideal has a basis consisting of two matrix elements in the right column, $\{N_3,(e_1N_3)\}$.  Again eight basis multivectors are projected onto them with real or imaginary coefficients.  Now they are all different.  But again there is a two-fold degeneracy in these projections, because we don't know if the prototype was positive, or negative.

From a unit cube representation one can see, that positive and negative Hilbert projections are complementary spinor images (shadows) of the same object.  To reconstruct the prototype unambiguously, it is enough to combine them in a strict sum.

\subsection{Hilbert projections in terms of rotations}

In a given frame of reference sometimes it is convenient to express a unit vector $a$ as a rotated (geometrically excited) state of a basis vector $e_3$:
\begin{equation}\label{22}
    a = Q(a)e_3\widetilde{Q}(a)
\end{equation}
Using $e_3 = P_3 - N_3$, one can change it into a difference of two complementary paravectors:
\begin{equation}\label{23}
    a = Q(a)P_3\widetilde{Q}(a) - Q(a)N_3\widetilde{Q}(a) = P(a) - N(a),
\end{equation}
where $P(a) = \frac{1}{2}(e_0 + a)$ and $N(a) = \frac{1}{2}(e_0 - a)$. Each paravector is an ordered composition of contravariant and covariant conjugated spinors. So (\ref{23}) can be written in the form
\begin{equation}\label{24}
    a = Q(a)P_3[Q(a)P_3]^\sim - Q(a)N_3[Q(a)N_3]^\sim = \psi_+(a)\widetilde{\psi}_+(a) - \psi_-(a)\widetilde{\psi}_-(a).
\end{equation}
In quantum mechanics positive spinors have special Dirac's notations --- bra- and ket- vectors:
\begin{equation}\label{25}
    \ket{a} = \psi_+(a) = Q(a)P_3 = \alpha P_3 + \beta (e_1P_3) =
    \left[ \begin{array}{cc}
    \alpha & 0 \\
    \beta & 0
    \end{array}\right];
\end{equation}
\begin{equation}\label{26}
    \bra{a} = \widetilde{\psi}_+(a) = P_3\widetilde{Q}(a) = \alpha^* P_3 + \beta^* (P_3e_1) =
    \left[ \begin{array}{cc}
    \alpha^* & \beta^* \\
    0 & 0
    \end{array}\right].
\end{equation}
The same Dirac's notations are the blinders for negative spinors:
\begin{equation}\label{27}
    \psi_-(a) = Q(a)N_3 = \alpha^* N_3 - \beta^* (e_1N_3) =
    \left[ \begin{array}{cc}
    0 & -\beta^* \\
    0 & \alpha^*
    \end{array}\right];
\end{equation}
\begin{equation}\label{28}
    \widetilde{\psi}_-(a) = N_3\widetilde{Q}(a) = \alpha N_3 - \beta (N_3e_1) =
    \left[ \begin{array}{cc}
    0 & 0 \\
    -\beta & \alpha
    \end{array}\right].
\end{equation}
They are never used in quantum mechanics. One of the reasons is their incompatibility with its probabilistic interpretation. So not to lose information we prefer to avoid bra- and ket- notations and probability terms in geometric algebra.

\begin{figure}[h]
\centering
\includegraphics[height=8cm]{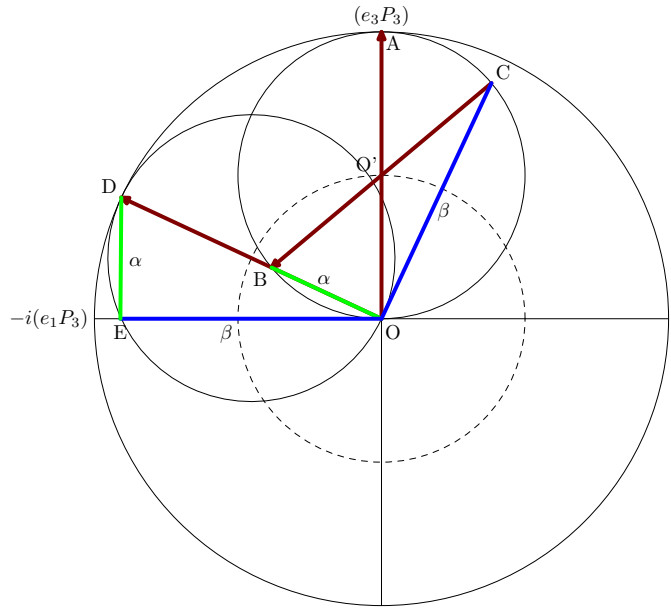}
\parbox[t]{0.9\textwidth}{
  \caption{A unit vector $a$ in Cartesian frame of reference and its contravariant spinor image $\alpha P_3 - i\beta (e_1P_3)$ in a positive Hilbert paravector space for a unit vector $e_3$.}\label{f:2}}
\end{figure}

There is a plenty of ways to get $a$ from $e_3$ in a simple rotation.  They differ from each other in rotation angle and axis values, which are coded in Cayley--Klein (or Euler--Rodrigues) parameters inside the first information layer.  Usually the shortest angle of rotation is chosen.  In this case as $e_3$ and $a$ are in the same plane of rotation, the axis of rotation is in $e_{12}$ plane, $\alpha$ is a real number, 4D Hilbert space is reduced to a 3D one, and the correspondence between spatial orientation of vector $a$ in Euclidean space and its positive contravariant spinor image in Hilbert space can be depictured in the same drawing \cite{tar_04}. A simple example of such 2D drawing is presented in Fig.\ref{f:2}.
Here a unit vector \emph{OA} is rotated anti-clockwise around $e_1$ axis, pinned to its center \emph{O'}, through an 130$^\circ$ angle into a \emph{CB} position. The small circle \emph{O'} with unit diameter corresponds to Euclidian space. It is rolled over the inner side of the big circle \emph{O} with a unit radius, corresponding to a positive Hilbert space of contravariant spinors for the selected direction $e_3$. When it is rolled through the arc of 65$^\circ$, \emph{CO} coincides with \emph{OE}, \emph{OB} coincides with \emph{ED}, \emph{CB} coincides with \emph{OD}. In this position Cayley--Klein parameters $\alpha$ and $\beta$ have the sense of directional cosines for Hilbert frame of reference. They are real in this example. Note, that in spinor representation, 1/2 is a scaling factor for the angle of rotation, not for the length of the vector.

In this 2D example the initial unit Euclidian vector \emph{OA} (small circle) coincides with its spinor image --- the Hilbert vector \emph{OA} (big circle), and the rotated Euclidian vector \emph{CB} (small circle) has quite another spinor image --- the Hilbert vector \emph{OD} (big circle). Note, that in general, vectors of Euclidian space are not vectors in Hilbert space and vice-versa. This example is of interest, because this very spinor image is used as a qubit in quantum information theory. There are attempts \cite{Nielsen_2002} to insert it in a Bloch or in a Poincare sphere, but these are just 3D analogs of the small Euclidian circle used in this example. Spinors and qubits live in 4D Hilbert spaces, which in general cannot be visualized in 3D Euclidian space.

In geometric algebra it is clear, that spinor images of a unit vector $a$ cannot be normed to a single real number. In one order (an inner product) the composition of contravariant and covariant shadows gives the the initial paravector for positive spinors:
\begin{equation}\label{29}
    \bra{a}a\rangle = P_3Q(a)\widetilde{Q}(a)P_3 = P_3 = \left[\begin{array}{cc}
    1 & 0 \\
    0 & 0
    \end{array}\right],
\end{equation}
and for negative spinors:
\begin{equation}\label{30}
    N_3Q(a)\widetilde{Q}(a)N_3 = N_3 = \left[\begin{array}{cc}
    0 & 0 \\
    0 & 1
    \end{array}\right].
\end{equation}
Their difference, $P_3 - N_3 = e_3$, gives the initial state of the rotated vector as $e_3$.  The condition $\alpha\alpha^* + \beta\beta^* = 1$ is an intrinsic part of definition for Cayley--Klein parameters in geometric algebra.

In the reversed order (an outer product) it gives the final paravector for positive spinors:
\begin{equation}\label{31}
    \ket{a}\bra{a} = Q(a)P_3P_3\widetilde{Q}(a) = P(a) = \left[\begin{array}{cc}
    \alpha\alpha^* & \alpha\beta^* \\
    \alpha^*\beta & \beta\beta^*
    \end{array}\right],
\end{equation}
and for negative spinors:
\begin{equation}\label{32}
    Q(a)N_3N_3\widetilde{Q}(a) = N(a) = \left[\begin{array}{cc}
    \beta\beta^* & -\alpha\beta^* \\
    -\alpha^*\beta & \alpha\alpha^*
    \end{array}\right].
\end{equation}
Their difference gives the final state $a$ of the rotated vector in terms of Cayley--Klein parameters:
\begin{equation}\label{33}
    a = P(a) - N(a) = \left[\begin{array}{cc}
    a_3 & a_1-ia_2 \\
    a_1+ia_2 & -a_3
    \end{array}\right] = \left[\begin{array}{cc}
    \alpha\alpha^* - \beta\beta^* & 2\alpha\beta^* \\
    2\alpha^*\beta & \beta\beta^* - \alpha\alpha^*
    \end{array}\right].
\end{equation}

So if a physical object which we want to use in information processing can be described as a vector of a 3D Euclidian space, and its states are associated with spatial orientations of this vector, then all information, we are interested in, is contained in Cayley--Klein or Euler--Rodrigues parameters inside the first information layer.  They have four real values, which are usually not conserved in time, and are not available for direct measurements.  There are no detectors for them in Hilbert spaces, but they can be measured in their compositions in 3D Euclidian space as stationary or rotating vector components, like in NMR pulsed experiments \cite{tar_02}.

\section{UNITARY OPERATIONS}

Unitary operations in geometric algebra are reflections and rotations. Their main feature is that, being applied to
any set of vectors, they keep all vector lengths and all angles among them invariant.  There are no deformations for any geometric object, except inversions.  So they are good for operation over composite geometric objects and for parallel operations over distributions of simple ones.  In our 8D algebraic space they can be used for information processing inside both information layers.

\subsection{Reflections}

There are three kinds of reflections in a 3D Euclidian space: reflection in a point, reflection in a line, and reflection in a plane.

Reflection in a point is a displacement of the frame origin in that point and spatial inversion of all vectors and their odd products in the new frame origin.  So all the vectors and trivectors change their signs.  Reflection in the origin of the frame of reference is marked with an overline. It is one of two main conjugation operations in geometric algebra. Structure elements (\ref{4}), joined with the unit cube main diagonals, are conjugated in that way.

Reflection in a line changes signs for all vector components, perpendicular to that line.  All components, collinear with that line conserve their signs.  For example, a reflection of vector (\ref{7}) in the line $e_1$ gives
\begin{equation}\label{34}
    e_1ae_1 = a_1e_1 - a_2e_2 - a_3e_3.
\end{equation}
It is a key to find reflection-in-line connections among 8D structure elements (\ref{4}). For example,
$A$ reflected in $e_1$ is $\overline{B}$, $A$ reflected in $e_2$ is $\overline{C}$, $A$ reflected in $e_3$ is $\overline{D}$, $B$ reflected in $e_3$ is $C$.

Reflection in a plane changes signs for all vector components, perpendicular to that plane. For example, a reflection of vector (\ref{7}) in the plane $e_{23}$ gives
\begin{equation}\label{35}
    e_{23}ae_{32} = -a_1e_1 + a_2e_2 + a_3e_3.
\end{equation}
Then for 8D structure elements (\ref{4}) one can get, for example, $A$ reflected in $e_{23}$ is $B$, $A$ reflected in $e_{13}$ is $C$, $A$ reflected in $e_{12}$ is $D$, $B$ reflected in $e_{12}$ is $\overline{C}$, etc.

So all structure elements (\ref{4}) are connected with each other through unitary operations of reflections in 3D Euclidian space.

Reflections in planes and lines, described by basis multivectors, change images only in the second information layer. Information images inside the first layer remain unchanged.  Reflections in arbitrary planes and lines change information images inside both layers.

\subsection{Rotations}

Spatial rotations are noncommutative operations, which cannot be described in single real or complex numbers. They need at least four real numbers or two complex numbers for each simple rotation.  Best of all they are described in terms of unitary quaternions.  The main sequence of their noncommutativity is that each simple rotation cannot be decomposed into superposition of simultaneous rotations in noncollinear planes or around noncollinear axes. There is no superposition principle for noncollinear components of a rotation axis.  A good example is forced rotations with nonzero offset in pulsed NMR experiments \cite{tar_02}.

In geometric information processing there is a practise to decompose a single simple rotation into a sequence of other simple rotations around noncollinear axes.  It can be done in a plenty of ways.  The main idea is to simplify some description in theory or to gain some advantages in excitation of inhomogeneously broadened mesomorphic structures in practise \cite{tar_04}.  The price is a significant and fast increase of duration for any such composite operation with each new discrete rotation element inside it.

Simple spatial rotations and their sequences are the main tools for geometric information processing inside the first information layer. Unitary quaternions, used to describe them, can be treated as universal 4D information units, independent from the physical objects to be rotated. In that sense they are similar to bits of information, but they live in 4D space and need more complicated logic to be operated with.

\section{COMPARISON WITH QUANTUM INFORMATION}

\subsection{Geometric qubit}

Using additive decomposition of the unit scalar into $P_3$ and $N_3$ paravectors, one can decompose unitary quaternion $Q(a)$ into a direct sum of a positive contravariant spinor, $Q(a)P_3$, and a negative contravariant spinor, $Q(a)N_3$:
\begin{equation}\label{36}
    Q(a) = Q(a)(P_3 + N_3) = Q(a)P_3 + Q(a)N_3.
\end{equation}
Positive spinor
\begin{equation}\label{37}
    Q(a)P_3 = \alpha P_3 + \beta (e_1P_3) =
    \left[ \begin{array}{cc}
      \alpha & 0 \\
      \beta & 0 \\
    \end{array} \right] =
    \alpha\,\includegraphics[height=1cm]{vers.31} +
    \beta\,\includegraphics[height=1cm]{vers.32}
\end{equation}
is a geometric analog of the object which is used in quantum mechanics as a wave function or a qubit. We shall call it geometric qubit here.  In terms of Euler--Rodrigues parameters it has a form
\begin{equation}\label{38}
    Q(a)P_3 = \rho \, \includegraphics[height=1cm]{vers.31} +
    \nu \, \includegraphics[height=1cm]{vers.33} +
    \mu \, \includegraphics[height=1cm]{vers.34} +
    \lambda \, \includegraphics[height=1cm]{vers.35}
\end{equation}
or
\begin{eqnarray}%
    Q(a)P_3 &=& (\rho - \nu - \mu - \lambda) \,
    \includegraphics[height=1cm]{vers.36} +
    (\rho + \nu + \mu - \lambda) \, \label{10a}
    \includegraphics[height=1cm]{vers.37} + \label{39} \\
\nonumber  &+& (\rho + \nu - \mu + \lambda) \,
    \includegraphics[height=.8cm]{vers.38} +
    (\rho - \nu + \mu + \lambda) \,
    \includegraphics[height=.8cm]{vers.39}.
\end{eqnarray}
One can see that this is a 4D object.  Its complementary counterpart is
\begin{equation}\label{40}
    Q(a)N_3 = -\beta^*(e_1N_3) + \alpha^* N_3 =
    \left[ \begin{array}{cc}
      0 & -\beta^* \\
      0 & \alpha^* \\
    \end{array} \right] =
    -\beta^*\,\includegraphics[height=1cm]{vers.40} +
    \alpha^*\,\includegraphics[height=1cm]{vers.41}
\end{equation}
In terms of Euler--Rodrigues parameters it has a form
\begin{equation}\label{41}
    Q(a)N_3 = \rho \, \includegraphics[height=1cm]{vers.41} +
    \nu \, \includegraphics[height=1cm]{vers.42} +
    \mu \, \includegraphics[height=1cm]{vers.43} +
    \lambda \, \includegraphics[height=1cm]{vers.44}
\end{equation}
or
\begin{eqnarray}
    Q(a)N_3 &=& (\rho - \nu - \mu - \lambda) \,
    \includegraphics[height=1cm]{vers.45} +
    (\rho + \nu + \mu - \lambda) \,
    \includegraphics[height=1cm]{vers.46} + \label{11a} \label{42} \\
\nonumber    &+& (\rho + \nu - \mu + \lambda) \,
    \includegraphics[height=.8cm]{vers.47} +
    (\rho - \nu + \mu + \lambda) \,
    \includegraphics[height=.8cm]{vers.48}.
\end{eqnarray}

Positive spinor, $Q(a)P_3$, can be treated either as an abstract mathematical quantity --- a part of a quaternion, --- describing an operation of a simple rotation, or as a result of its application to the vector $e_3$ or to the scalar $e_0$.

As mathematical quantity it is used to design calculation algorithms. Simple rotations are composed in sequences through ordered products of corresponding unitary quaternions. And there are two complementary spinors inside the same resulting unitary quaternion, associated with the same geometric qubit.

As a result of a unitary operation spinor changes its image inside the first information layer.  But for a single Hilbert projection one cannot distinguish between a rotated vector $e_3$ and a rotated scalar $e_0$.  In geometry there is a difference.  Vector changes its orientation in Euclidian space, and hence its state or image inside the first information layer.  Scalar is invariant to any rotation.  So there will be no changes in the first information layer, and the equation of identity will always be true.  To prevent the loss of information inside the first layer one needs to use as positive, and negative spinor images for any qubit.

In geometric algebra not only vectors, but also bivectors and their cluster combinations with all other elements of $\mathcal{G}_{3,0}$ algebra can be rotated. Sometimes they have or gain explicit vector components, which could interfere with calculation results designed only for vectors. Such unpredictable gains and losses of information are inadmissible for large-scale calculations, especially for quantum ones.  So it is a good idea to treat them geometrically as well.

In quantum information processing real or complex numbers inside the first layer are treated as probabilities or their amplitudes and are thrown away in averaging operations.  Hence all calculations seem to be performed only inside the second information layer, over structure elements (\ref{4}) of a computational basis, that is over geometric byte structure. If it is so, then it is a pure mathematics, and where is physics?

\subsection{Hadamard gate}

Usually as a qubit and its geometric analog is written in a form (\ref{37}). Its computational basis consists of two paravectors: $P_3$ and $(e_1P_3)$.  Hadamard gate is used to change it to positive and negative superpositions of both.  Due to two-fold degeneracy in Hilbert projection results, there are two ways to do it.  If we assume, that $P_3 = e_3P_3$, then we could sum or subtract unit vectors $e_1$ and $e_3$ in Cartesian frame of reference, norm them to $e_0$, Hilbert project them with $P_3$, and use them as new basis elements, $\{ \frac{1}{\sqrt{2}}(e_1+e_3)P_3,\frac{1}{\sqrt{2}}(e_1-e_3)P_3 \}$.  This way is used in quantum information processing \cite{Nielsen_2002}.

The other way is to assume, that $P_3 = e_0P_3$, then to decompose $e_0$ and $e_1$ into their clusters of likeness and distinction, $e_0 = P_1 + N_1$ and $e_1 = P_1 - N_1$. After that one can decompose basis paravectors, $P_3$ and $(e_1P_3)$, into $e_0P_3 = P_1P_3 + N_1P_3$ and $(e_1P_3) = P_1P_3 - N_1P_3$, and to regroup them:
\begin{eqnarray}
\nonumber    Q(a)P_3 &=& \alpha P_3 + \beta (e_1P_3) = \alpha (P_1P_3 + N_1P_3) + \beta(P_1P_3 - N_1P_3) = \\
   &=& (\alpha + \beta)(P_1P_3) + (\alpha - \beta)(N_1P_3) =
(\alpha + \beta)(A + C) + (\alpha - \beta)(B + \overline{D}).  \label{43}
\end{eqnarray}
This is an ordinary regrouping operation for structure elements (\ref{4}) inside the second information layer.  Note, that 1/2 factor is attributed now to $P_1$ and $N_1$ to make them idempotent ones.  This way is typical for geometric information processing.

\subsection{NOT gates}

Both in quantum and in geometric information processing, reflections in lines, associated with basis vectors of Cartesian frame of reference, are used to implement NOT operations.  In geometric algebra it is trivial to use reflection in $e_1$ as NOT operation for any unit vector in (\ref{22}) form.  Its positive contravariant Hilbert projection has a geometrical qubit (\ref{37}) form.  Multiplication by $e_1$ from the left changes $P_3$ to $(e_1P_3)$, and $(e_1P_3)$ to $P_3$, because $e_1$ is a unit vector, $e_1^2 = e_0$.  Information inside the first information layer remains unchanged.  These operations are applied only to the second information layer.

Note that in geometric algebra $P_3$ and $N_3$ are associated with ``spin-up'' and ``spin-down'' states, correspondingly, rather than  basis paravectors $P_3$ and $(e_1P_3)$, which are just positive contravariant Hilbert projections for perpendicular basis vectors $e_3$ and $e_1$ of Cartesian frame of reference, respectively. Both of them are in the same positive Hilbert space, defined by $P_3$ paravector.

\subsection{Phase and coherence}

Inside the first information layer each state of a geometric qubit (\ref{37}) is fully described by two complementary complex Cayley--Klein parameters, $\alpha$ and $\beta$. They are related with equations (\ref{13}), which are more strict and general, then in eq. (1.3) of quantum information \cite{Nielsen_2002}. Although they are complex numbers and can be represented in a ``module--phase'' form, it is impossible to compare them by their phase, because their phases are defined in quite different planes.

In quantum information \cite{Nielsen_2002} basis paravectors, $P_3$ and $(e_1P_3)$, are treated as the states, $|0\rangle$ and $|1\rangle$, of a qubit, and eq. (\ref{37}) --- as their superposition, with probability amplitudes $\alpha$ and $\beta$. The latter are used as a measure of coherence between $|0\rangle$ and $|1\rangle$ states.

In geometric algebra the picture is quite different. Basis paravectors, $P_3$ and $(e_1P_3)$, are not the states, but rather Hilbert projections of Cartesian frame of reference, which is always a stable and unchangeable object. They are always coherent as parts of the same frame of reference. They have nothing to do with a particular physical or geometric object we are going to describe in that frame of reference.  Cayley--Klein parameters are coherent, but not through their phase relations.  They are coherent in the sense, that they describe one of four Hilbert projections for a particular simple rotation.  One can feel some decoherence only for a rotation in a spatially unstable plane or around some axis with unstable spatial orientation.

Geometric algebra is not blind to phase. Each phase is defined in its plane of rotation.  And there are plenty of such planes in its 8D algebraic space. It contains as operators, and operands. In geometric algebra operators can be expressed in terms of operands and vise-versa. But in this algebra there is no place for such monsters as phase operator $S = P_3 + iN_3$ \cite{Nielsen_2002}.

\section{DISCUSSION}

The described approach is of interest for perfection of information processing technologies in a number of ways.
Its main idea is very close to ideas of R.B. Fuller and N. Tesla: each simplicity is a pretty organized complexity.
We tried to extend calculation inside geometric Clifford algebras onto some new sound 4D paravector clusters with some new algebraic properties. We used these clusters to define geometric bits and to organize them into geometric byte structure. It gave us the possibility to define two layers of information: a geometric and an algebraic ones. To deal inside them with only real numbers we changed matrix representations onto an oriented unit cube ones. We described two ways to project algebraic and geometric objects onto subspaces of less dimensions and gave some examples to apply them in practise.

We tried to use mathematical similarities in geometric algebra formulas with those in quantum mechanics to compare our approach with that of quantum information processing. We hope that geometric approach will help to avoid some information ambiguities in future technologies of quantum computations.

The possibility to work only with real numbers is very good for ordinary classical computers. Now they are smart enough to process eight flows of information in parallel. There are two layers of information to work with in each flow. If information processing inside the first layer is reduced to manipulations with separate symbols, information processing inside the second layer can be associated with fast reading technologies. For eight independent flows of a discrete or analog information inside the first layer there is a possibility to mix them coherently with operations inside the second layer and to separate them afterwards in constructive interference processes without any claims on their coherence or multiplexing in temporal or frequency domains. For coherent flows of information, for example ones describing coordinates of some moving (rotated) 8D geometric object, there is a unique possibility to simulate the behavior of such an object on a classical computer.

This article is one of the first steps in this direction.  We hope that some of our ideas could be helpful for our colleagues as in Clifford algebras applications and in quantum information processing area.

\end{document}